# Surface oxidation and thermoelectric properties of indium-doped tin telluride nanowires

Zhen Li,[a] Enzhi Xu,[a] Yaroslav Losovyj,[b] Nan Li,[c] Aiping Chen,[c] Brian Swartzentruber,[d] Nikolai Sinitsyn,[e] Jinkyoung Yoo,[c] Quanxi Jia[c,f] and Shixiong Zhang*[a]

a. Department of Physics, Indiana University, Bloomington, IN 47405, USAE-mail: sxzhang@indiana.edu
b. Department of Chemistry, Indiana University, Bloomington, IN 47405, USA
c. Center for Integrated Nanotechnologies, Los Alamos National Laboratory, Los Alamos, NM 87545, USA
d. Center for Integrated Nanotechnologies, Sandia National Laboratories, Albuquerque, NM 87185, USA
e. Theoretical Division, Los Alamos National Laboratory, Los Alamos, NM 87545, USA
f. Department of Materials Design and Innovation, University at Buffalo – The State University of New York, Buffalo, NY 14260, USA

**Abstract**

The recent discovery of excellent thermoelectric properties and topological surface states in SnTe-based compounds has attracted extensive attention in various research areas. Indium doped SnTe is of particular interest because, depending on the doping level, it can either generate resonant states in the bulk valence band leading to enhanced thermoelectric properties, or induce superconductivity that coexists with topological states. Here we report on the vapor deposition of In-doped SnTe nanowires and the study of their surface oxidation and thermoelectric properties. The nanowire growth is assisted by Au catalysts, and their morphologies vary as a function of substrate position and temperature. Transmission electron microscopy characterization reveals the formation of amorphous surface in single crystalline nanowires. X-ray photoelectron spectroscopy studies suggest that the nanowire surface is composed of $In_2O_3$, $SnO_2$, Te and $TeO_2$ which can be readily removed by argon ion sputtering. Exposure of the cleaned nanowires to atmosphere yields rapid oxidation of the surface within only one minute. Characterizations of electrical conductivity σ, thermopower *S*, and thermal conductivity κ were performed on the same In-doped nanowire which shows suppressed *σ* and *κ* but enhanced *S* yielding an improved thermoelectric figure of merit ZT than the undoped SnTe.

**Introduction**

Tin telluride (SnTe)-based narrow band-gap semiconductors have recently triggered growing interest owning to the demonstration of excellent thermoelectric (TE) properties [1-10] and the discovery of topological surface states.[11-20] The TE figure of merit of a material is defined by a dimensionless parameter: $ZT = \frac{S^2 \sigma T}{\kappa}$, where $S$ is the thermopower or Seebeck coefficient, $\sigma$ the electrical conductivity and $\kappa$ the thermal conductivity. While the pristine bulk SnTe crystal is not compelling because of its low thermopower and high thermal conductivity,[21] doped and alloyed SnTe exhibits significantly enhanced TE performance.[1, 2, 22-27] In particular, less than 2% indium-doping can induce resonant states in the valence band of SnTe, which dramatically improve its thermopower and hence the ZT.[1] The peak ZT achieved in In-doped SnTe is as high as 1.1 at 873 K, suggesting its great potential to replace lead chalcogenides for applications in waste heater recovery.[1] Indium doping in SnTe can also lead to superconductivity coexisting with topological surface states[28-33] with potential applications for topological quantum computing.[34]

In comparison to bulk crystals, nanostructures such as nanowires provide a unique platform to improve thermoelectric performance and to exploit toplogical states. Indeed, nanoscale size can enhance phonon-surface scattering, which could significantly suppress thermal conductivity and hence boost the figure of merit ZT.[35-49] From a topological state point of view, the large surface-area-to-volume ratio of nanostructures magnifies contributions from surface states over bulk states of the material.[50-55] The topological surface could have enhanced thermopower due to its unique energy dispersion and strongly energy-dependent relaxation time.[49, 56]

In spite of great promise in various aspects, the surfaces of SnTe-based compounds are often prone to contamination or oxidation [57-59]. This creates an essential challenge to directly probing topological states of pristine nanostructures using surface sensitive techniques such as scanning tunneling microscopy (STM) and nano angle-resolved photoemission spectroscopy (nano-ARPES), because of the unavoidable exposure to air during sample transfer from the growth reactor to the measurement chamber. Furthermore, it is unclear if the surface oxidation would impact the materials properties. Therefore, a thorough investigation on the surface oxidation of nanostructured SnTe is essential, and a search for more effective surface cleaning methods is imperative.

In this work, we carried out surface oxidation, cleaning and thermoelectric studies of In-doped SnTe nanowires grown by chemical vapor deposition. While the growth of undoped SnTe nanowires has been well studied over the past few years,[50-52] chemical doping, which may enhance physical properties and introduce new functionality,[60] are scarcely reported in SnTe nanowires.[61, 62] In contrast to the catalyst-free growth of nanoplates[55, 63] and thin films,[64] Au nanoparticles were employed to catalyze the growth of nanowires. The surface of In-doped SnTe was characterized by transmission electron microscopy (TEM) and X-ray photoelectron spectroscopy (XPS) studies, which revealed the formation of $In_2O_3$, $SnO_2$, Te and $TeO_2$ due to rapid oxidation in exposure to atmosphere. The surface oxides can be readily removed by gentle argon ion sputtering, yielding a potential approach to cleaning the surface of *ex-situ* grown nanostructures for surface sensitive measurements such as ARPES and scanning tunneling microscopy (STM). Characterizations of electrical conductivity σ, thermopower S, and thermal conductivity κ were performed on the same In-doped SnTe nanowire which shows suppressed σ and κ but enhanced S, rendering an improved thermoelectric figure of merit ZT in comparison to the undoped SnTe.

**Experimental Details**

**1. Synthesis**

In-doped SnTe nanowires were synthesized inside a three-zone tube furnace (Figure 1a) via a process similar to the growth of undoped SnTe nanowires.[50-52] Tin telluride powder (Alfa Aesar, purity 99.999%) was placed in a combustion boat located in the center of the tube furnace, indium powder (Alfa Aesar, purity 99.99%) and tellurium powder (Alfa Aesar, purity 99.999%) were placed in the upstream position 8 inches away from the center. Three ~3-inch long silicon substrates coated with gold colloids (Ted Pella, diameter 80 nm) were placed adjacent to each other in the downstream position from 4 inches to 13 inches away from the center of the tube furnace. Ultra-high purity argon, as a carrier gas, was introduced and kept at a flow rate of 10 sccm with a pressure maintained at 30 Torr. During the growth of nanowires, temperatures of three zones in the tube furnace were kept at 480/600/475 °C (set points) for 5 hours (see the main text and Figure S1 for the actual growth temperatures measured using an external thermocouple). After the growth was complete, the heater was turned off and the system was cooled to room temperature over about 5 hours.

## 2. Characterization

The morphology of nanostructures was characterized by a scanning electron microscope (SEM, Quanta FEI). The crystal structure and orientation were characterized by TEM (FEI Tecnai F30). Raman measurements were performed in a confocal Raman microscope system (Renishaw inVia) using a 532 nm excitation laser under ambient environment. The chemical composition was characterized by Energy-dispersive X-ray Spectroscopy (EDX) in both SEM and TEM. The nanowires were examined by X-ray diffraction (XRD) (Rigaku Ultima III), using Cu-K$_\alpha$ radiation ($\lambda$=1.5406 Å). XPS measurements were conducted using a PHI *Versa Probe* II instrument. X-ray beam with a power of 50 W at 15 kV and a spot size of 200 μm was generated by a monochromatic Al $k_\alpha$ source. The calibration of instrument work function was carried out using a standard procedure described previously.[65] For all samples, neutralization was provided by using the PHI dual charge compensation system. XPS spectra were taken at the energy step of 0.1 eV using SmartSoft–XPS v2.3.1 at pass energies of 46.95 eV for Te 3d, 23.5 eV for C 1s, 11.75 eV for Sn 3d, and 93.9 eV for In 3d transitions. The spectra were calibrated using either highly oriented pyrolytic graphite or carbon 1s peak and were further analyzed in PHI MultiPack v9.0 software. Peaks were fitted using GL line shapes with 80% Gaussians and 20% Lorentzians, along with a Shirley background.

## 3. Device fabrication and electrical measurements

The thermoelectric devices were fabricated and measured using the approaches reported earlier.[66-69] The thermoelectric platform with four electrodes and a resistive heater was fabricated on top of the SiN$_x$/SiO$_2$/Si substrate by e-beam lithography (EBL) followed by metal deposition of ~ 10 nm Ti / 90 nm Au. A trench was etched in the middle region of the platform to limit heat transfer on the surface of the substrate. An In-doped SnTe nanowire was picked up by a nanomanipulator, and then placed on the platform. EBL and metal deposition (~ 10 nm Ti /290 nm Au) was performed again to connect the nanowire with the electrodes on the platform. Right before metal deposition, the sample was cleaned by oxygen plasma and then soaked in 1% HCl for ~80 seconds to remove remaining e-beam resist and oxides on the surface of the nanowire. The electrical conductivity was determined by a standard four-probe measurement of resistance. The thermopower was measured by applying a current through the heater, creating a temperature gradient along the nanowire. The thermoelectric voltage V$_{th}$ between B and C were measured. Electrodes B and C also serve as thermometers by measuring their resistances, and the temperature difference ΔT between B and C was obtained. Thermopower of the device was then calculated by $S = V_{th}/\Delta T$, and the thermopower of the nanowire was determined by subtracting contribution from the Au electrodes. The thermal conductivity was measured using a self-heating method: by applying a large current through electrodes A and D, a voltage between B and C was measured and the resistance was calculated.[66-70] The average temperature increase of the nanowire $\Delta T_M$ was then determined from the resistance change (a linear temperature dependence is assumed in a 10 K interval). The thermal conductivity can be calculated based on $\kappa = I^2 R l/(12 \Delta T_M A)$, where *I* is the current, *R* is the resistance, *l* is the channel length, and *A* is the cross section of the nanowire.[66-70]

**Results and Discussion**

Figure 1a shows a schematic of the chemical vapor deposition system for nanowire synthesis. Using Au as catalyst, indium, tellurium and tin telluride as precursors, we achieved the growth of In-doped SnTe nanowires with a variety of morphologies. A high density of straight and smooth nanowires with a typical width of 100 – 200 nm and length of about 5 – 8 μm were observed on the first substrate (closest to the precursors) in a temperature range of 605 ˚C ~ 525 ˚C (Figure 1b). On the second substrate (further away from the precursors) where the temperature is 525 ˚C ~ 487 ˚C, we observed large, smooth nanobeams with a typical width of about 1 μm and length of 20 – 50 μm (Figure 1c). When the temperature is further reduced on the third substrate (farthest from the precursors) to 487 ˚C ~ 359 ˚C, 'twisted' nanowires are formed (Figure 1d). The typical width of the 'twisted' nanowires varies from about 100 nm to 300 nm and the length is 1 – 3 μm. The insets of Figures 1b-d are the corresponding high-magnification SEM images taken near the tip of the nanowires. Au-Sn-Te alloy particle is observed on the tip of these nanowires/nanobeams, indicating a vapor-liquid-solid growth mechanism (Figure S2, Supplementary Information). The nanowire diameter is usually larger than the particle, suggesting direct vapor-solid deposition on the side wall of the nanowire. Along with the nanowires are a high density of crystals whose size ranges from nanometers to micrometers.

To understand their structures and to check phase purity, XRD $\theta - 2\theta$ scan was performed on the nanowires on the first, second and third substrate as well as on a bare silicon substrate for comparison. As shown in Figure 2, all diffraction peaks except for Si (400) from three samples correspond to a rock-salt face-centered-cubic phase of SnTe with no other impurity phases within the detection limit of XRD. The multiple peaks from different planes of the cubic phase is consistent with the fact that the nanowires and crystals lay randomly on the substrate. The (200) peak has the highest intensity, which agrees qualitatively with the previous results of polycrystalline samples.[1] The intensity of the peaks from the third substrate is relatively low, which is possibly due to the lower structural quality of the nanowires (as shown below).

We further conducted TEM and Raman measurements to characterize the structure of individual nanowires. Figure 3a shows a TEM image taken on the body of a typical nanowire from the first substrate, the nanowire surface is smooth and the width of the nanowire is about 104 nm. The selected area electron diffraction (SAED) pattern is shown in the inset, which suggests the single crystallinity of the nanowire and the growth direction along [100]. A high resolution TEM (HRTEM) image is shown in Figure 3b and its inset, the d-spacing between crystal planes perpendicular to the growth direction is measured to be about 0.32 nm, corresponding to the spacing of two adjacent (200) planes. Fast Fourier transform (FFT) was carried out near the edge and the core of the nanowire body respectively. The edge of the nanowire body appears to be amorphous with a circular FFT pattern absent of most diffraction spots as shown in Figure 3c. Whereas FFT taken on the core has clear diffraction spots indicating the crystalline nature (Figure 3d). The distinction in crystallinity between the edge and the core of the nanowire is likely caused by surface oxidation of the compound (as further discussed below). Due to the large thickness of the nanowires from the second substrate, they are not transparent to electron beam in the TEM. However, the resemblance of morphologies between nanowires from the first substrate and the second substrate along with similar SnTe peaks from the XRD pattern implies that the nanowires from the second substrate are also single crystalline grown along [100]. A Raman spectrum taken on a typical nanowire shows two prominent peaks at ~ 124 cm$^{-1}$ and ~142 cm$^{-1}$ (Figure 3e). While all phonon modes are Raman inactive in a rock-salt crystal structure, these peaks were also observed in both undoped and Mn-doped SnTe bulk crystals, which was attributed to the presence of defects (e.g. Sn vacancies) or the breaking of symmetry on the surface.[71-73]

To characterize the compositions of nanowires, we carried out EDX measurements. As shown in Figure S3 (Supplementary Information), Sn and Te peaks are observed in the EDX spectra obtained by single point acquisition in SEM. Indium peaks are not observed, indicating that the dopant concentration is below the detection limit of EDX. The Sn:Te ratio varies slightly from one nanowire to another but is close to ~1 within the accuracy of the measurement. EDX mapping in Figure 3f shows uniform elemental distribution of Sn and Te in the core of the nanowire. Towards the surface, both Sn and Te signals drop and Te appears more deficient compared to Sn.

TEM and EDX mapping were performed on nanowires from the third substrate as well. As shown in Figure 4a, the nanowire is straight in the body but 'twisted' in the head. Figures 4b and c are the FFT patterns taken from the circled areas in Figure 4a. Multiple crystal orientations exist, suggesting a polycrystalline character of the nanowire. The EDX mapping (Figure 4d) shows relatively uniform elemental distribution of Sn and Te. The signals of Sn and Te again look weaker near the surface but the atomic ratio between Sn and Te is mostly unchanged.

From the TEM and EDX analysis, it can be summarized that the growth of single crystalline straight nanowires is favored in the high temperature zone near the precursors and the polycrystalline 'twisted' nanowires are preferred in the low temperature zone far away from the precursors (Detailed temperature profile in Figure S1). To investigate whether the temperature or the distance to the precursors plays a major role in determining the nanowire morphology, we carried out another growth with the same conditions as before but a temperature 50 °C lower in the third zone of the furnace where the three substrates were located. This time, nanowire morphologies are reproduced but shifted closer to the center of the furnace relative to the original growth, i.e. large and smooth nanobeams now appear on the first substrate (Figure S4a) instead of the second (Figure 1c), 'twisted' nanowires now are formed on the second substrate (Figure S4b) rather than the third (Figure 1d). Hence, we conclude that the temperature of the substrate is the prime factor in determining the nanowire morphologies, and a higher temperature yields single crystalline straight nanowires while a lower temperature gives rise to polycrystalline 'twisted' nanowires.

Next, we conducted XPS measurements to understand the nature of the amorphous surface and the Te deficiency in the nanowires grown on the first substrate. XPS spectra taken on a pristine sample (i.e. before sputtering as discussed below) reveal clearly the presence of oxides on the surface. A strong oxygen 1s peak was observed in Figure 5a, which is fitted by two components where one component with a higher binding energy corresponds to -2 state in $SiO_2$ from the substrate, and the other is related to -2 state in oxides from SnTe surface. The Sn 3d spectrum in Figure 5b shows two peaks at ~ 496.0 eV ($3d_{3/2}$) and ~ 487.6 eV ($3d_{5/2}$), indicating a +4 oxidation state in contrast to the nominal +2 in SnTe.[57-59] The Te 3d spectrum in Figure 5c has two components with binding energies of 583.9 eV ($3d_{3/2}$), 573.5 eV ($3d_{5/2}$) and 587.8 eV ($3d_{3/2}$), 577.4 eV ($3d_{5/2}$), respectively. The former corresponds to a 0 state,[74] and the latter is +4 as in $TeO_2$.[57-59] In contrast to the EDX characterization which has a lower sensitivity, two prominent In peaks are observed in Figure 5d and the binding energies are 453.7 eV ($3d_{3/2}$) and 446.0 eV ($3d_{5/2}$) which correspond to a +3 oxidation state.[75] These results suggest that the sample surface is oxidized and forms $In_2O_3$, $SnO_2$, Te and $TeO_2$. The atomic ratio determined based on the XPS spectra is In: Sn: Te=3.3:77.4:19.3.

XPS spectra taken at other spots on the first substrate show similar results (Figure S5 and Table S1, Supplementary Information). The large deviation of Sn:Te ratio from the nominal value in SnTe was previously observed in undoped SnTe bulk crystals[58] and thin films[57] as well, and is attributed to the different oxidation rates of the two elements.

XPS spectra were also taken on the second and the third substrates (Figures S6 and S7, Supplementary Information). On the second substrate, the Sn 3d and Te 3d core level spectra are qualitatively similar to those on the first substrate. The more symmetric oxygen peak in Figure S6a indicates that the component from $SiO_2$ is dominant over the contribution from the oxides of SnTe surface, which could be caused by the low density of nanostructures exposing larger area of $SiO_2$ to the X-ray beam (Figure 1c). In addition, no In peak was observed (Figure S6d), indicating that the indium concentration on the second substrate is below the detection limit. The atomic ratio determined from the XPS spectra taken at one of the three spots (Table S2, Supplementary Information) is Sn: Te=82.3:17.7, again suggesting a surface with excessive oxides of Sn. On the third substrate, for the oxygen peak (Figure S7a), the contribution from $SiO_2$ and from SnTe surface oxides are comparable, suggesting that large area of $SiO_2$ is covered by dense nanostructures, which also agrees well with SEM observation (Figure 1d). Furthermore, strong In 3d peaks were observed (Figure S7d), and quantitative analysis of a representative set of data shows an atomic ratio of In: Sn: Te=9.1:41.6:49.3 (Figure S7 and Table S3, Supplementary Information).

It is worth noting that the indium concentration detected by XPS does not show a monotonic dependence on the substrate location but instead decreases from the first substrate to the second and then increases in the third substrate. We believe, in contrast to the nanowire morphologies which solely depend on the temperature, the variation of indium concentration is related to both the temperature and the distance from the precursors to the substrates. On one hand, indium has high vapor pressure and it requires low temperature to form in a solid phase. As shown in Figure S1, the temperature decreases from the first substrate to the third, indium would form a solid phase more easily in the low temperature regions. On the other hand, a higher partial pressure of indium makes it easier to incorporate into a solid phase. In the furnace, the indium partial pressure decreases from the first substrate to the third since the distance between the precursors and substrates increases. Hence, the competition between the temperature effect and the effect of indium partial pressure leads to the non-monotonic dependence of indium concentration on the substrate location.

As discussed above, the presence of surface oxides is a critical challenge to study the topological states via surface sensitive techniques such as STM and nano-ARPES. Therefore, it is important to develop an effective approach to removing the surface oxides. We demonstrate here that argon sputtering can be used to achieve this purpose. Figure 5 compares the XPS spectra before and after sputtering. The sputtering process was performed for 1, 2, and 3 hours. After 1 hour of sputtering, the oxygen 1s peak is significantly reduced (Figure 5a). In the meantime, the $Sn^{4+}$ peaks disappeared, accompanied by the appearance of $Sn^{2+}$ peaks with lower binding energies of 494.1 eV and 485.6 eV (Figure 5b). The $Te^{4+}$ and Te peaks also disappeared, while the $Te^{2-}$ peaks became visible with binding energies of 583.2 eV and 572.7 eV (Figure 5c). This result suggests that the surface $SnO_2$, Te and $TeO_2$ were removed by sputtering. SEM images were taken after sputtering, the surfaces of the microcrystals and nanowires appear apparently etched (Figure S8, Supplementary Information). Upon the removal of surface $In_2O_3$, the In 3d peaks were shifted to lower binding energies and the peak intensities were reduced. The atomic ratio of In: Sn: Te after 1 hour of oxide removal becomes: 0.6:55.1:44.3, corresponding to an indium doping concertation of ~ 1%. The XPS spectra do not change notably under further sputtering, and the atomic ratio determined from the spectra does not show a strong variation as a function of sputtering time (Table 1). We therefore conclude that the sample composition is quite uniform in the non-oxidized region.

We further carried out re-oxidation experiments to demonstrate that surface oxidation occurs rapidly in atmospheric environment. Figures 6a-c compare the XPS spectra taken right after the sample was cleaned by sputtering for 30 minutes and after it was exposed to air for 1, 2 and 4 mins, respectively. Right after sputtering, the Sn and Te are mainly in +2 and -2 states, respectively. The slight asymmetry of the Te peaks is caused by a small amount (~15%) of residual Te due to incomplete cleaning. After only 1 minute of exposure to air, $Sn^{4+}$ peaks appear and dominate over the $Sn^{2+}$; the $Te^{4+}$ peaks are also observed in spite of a weak intensity. The oxygen 1s peak was significantly enhanced accordingly. The change in spectra is relatively small under further exposure for 2 and 4 minutes. To obtain a quantitative picture, we fitted the XPS spectra of Sn $3d_{5/2}$ using two components ($Sn^{4+}$ and $Sn^{2+}$) and Te $3d_{5/2}$ using three components ($Te^{4+}$, Te and $Te^{2-}$) and plot the $Sn^{4+}/Sn_{total}$ and $Te^{4+}/Te_{total}$ ratios as a function of exposure time in Figure 6d. It is clear that the oxidation occurs rapidly within 1 minute of exposure and then the oxidation rate decreases significantly. Also in Table 1, the atomic ratio determined from the spectra taken after 1 minute of exposure to atmosphere shows a sudden change compared to that obtained immediately after sputtering, but longer exposures of 2 and 4 minutes do not induce further significant changes in atomic ratio. This suggests that substantial oxidation occurs within 1 minute of exposure.

Lastly, we demonstrate the enhanced thermoelectric figure of merit of In-doped SnTe nanowires in comparison to the undoped sample by measuring the electrical conductivity, thermopower and thermal conductivity of the same nanowire (Figure 7a) on the first substrate. The detailed information of device fabrication and measurment is provided in the Experimental Details. It is extremely difficult to avoid contact with air during device fabrication process, so the nanowire being measured here is already oxidized. However, due to the amorphous and insulating nature of the thin oxidized surface, the measured thermoelectric properties should be mainly contributed from the single crystalline In-SnTe core. The electrical conductivity σ determined from a standard four-terminal measurement is ~$1.49 \times 10^5$ S $m^{-1}$ at 300 K, which is close to the conductivity of In-doped SnTe polycrystalline samples (~$2.0 - 4.3 \times 10^5$ S $m^{-1}$) [1] and nanoplates (~$1.4 \times 10^5$ S $m^{-1}$),[63] but is clearly lower than that of the undoped SnTe nanowires ($6 - 7 \times 10^5$ S $m^{-1}$).[67] The σ increases when the sample is cooled down to around 100 K and then starts to decrease, rendering a metal-semiconductor transition (Figure 7b) which is in contrast to the pure metallic behavior observed in the undoped SnTe.[67] The metallic behavior in SnTe is attributed to degenerate doping by native Sn vacancies which act as acceptors (or p-type dopants). The density of Sn vacancies is estimated to be on the order of $10^{20}$ $cm^{-3}$ (or 1%),[31, 67] comparable to the indium dopant concentration in the core of the wire. While indium is considered as a p-type dopant as well, it provides less holes than a Sn vacancy.[1] So when In atoms fill in the Sn vacancies, the hole density decreases. Another influence of indium doping is the introduction of disorder, which reduces hole mobility by enhancing carrier-impurity scattering. Both the reduction of hole density and mobility could be responsible for the suppression of electrical conductivity upon doping.

In contrast to its suppressed electrical conductivity, the In-doped SnTe nanowire shows enhanced thermopower S in comparison to the undoped SnTe. The temperature dependence of S is shown in Figure 7b. At room temperature, the thermopower is 59 µV K$^{-1}$, which is comparable with the values measured on the polycrystalline samples (~50 – 75 µV K$^{-1}$)[1] and is enhanced relative to the undoped SnTe nanowires (19 – 41 µV K$^{-1}$).[67] The positive sign of S confirms the p-type nature of the majority charge carriers. The thermopower decreases monotonically with the decrease of temperature and shows no bipolar effect.

The total thermal conductivity κ of the nanowire was measured using a self-heating method.[66-70] As shown in Figure 7c, the κ at 300 K is 2.24 W m$^{-1}$ K$^{-1}$, slightly lower than that of both the In-doped SnTe pollycrystal (~3.8 – 4.4 W m$^{-1}$ K$^{-1}$)[1] and the undoped SnTe nanowires (~4.2 W m$^{-1}$ K$^{-1}$).[67] The total thermal conductivity can be decomposed into two parts: electronic contribution and lattice contribution. The electronic thermal conductivity κ$_e$ is determined based on the Wiedemann-Franz law, i.e. $\kappa_e = L\sigma T$, where *L* is the Lorenz number. We calculated the Lorentz number using the same model for bulk SnTe in which contributions from both the light hole band and heavy hole band are included.[1] The calculated Lorenz number (e.g. 1.65 × 10$^{-8}$ V$^2$ K$^{-2}$ at 300 K) falls in the range of ~1.49 × 10$^{-8}$ V$^2$ K$^{-2}$ for nondegenerate semiconductors and ~2.44 × 10$^{-8}$ V$^2$ K$^{-2}$ for metals.[76] The lattice thermal conductivity κ$_L$ is then determined by subtracting the total thermal conductivity by the electronic thermal conductivity. Both κ$_e$ and κ$_L$ are plotted in Figure 7c, κ$_e$ increases as the temperature is raised, while κ$_L$ shows a non-monotonic behavior with a maximum value at 60K. The non-monotonic relationship between κ$_L$ and temperature may be due to the increase of phonon density upon warming at lower temperatures and the enhancement of umklapp phonon-phonon scattering at higher temperatures.[67] The ZT was finally calculated and plotted as a function of temperature in Figure 7d. At 300 K, the ZT of our nanowire is 0.07, which is comparable to that of In-doped SnTe polycrystalline samples (~ 0.08 – 0.09)[1] but higher than that of undoped SnTe nanowires (~0.0379).[67] The enhancement of ZT in the indium-doped nanowire should be attributed to both the improved thermopower and the suppressed thermal conductivity.

**Conclusions**

We presented the first growth of In-doped SnTe nanowires via chemical vapor deposition. The nanowires were grown under the assistance of Au catalysts, with various morphologies depending on the substrate growth temperature. The surface of the nanowires is amorphous contrary to the single crystalline core as revealed by transmission electron microscopy. The surface amorphous layer was further studied by X-ray photoelectron spectroscopy which suggested its composition to be $In_2O_3$, $SnO_2$, Te and $TeO_2$. Argon ion sputtering was found to effectively remove the amorphous layer. We also demonstrated that exposure of the cleaned nanowires to air within one minute led to rapid oxidation of the surface. Compared to undoped SnTe, an enhanced thermoelectric figure of merit ZT was obtained from our nanowires doped by ~1% of indium. The enhancement is attributed to the suppression of thermal conductivities and the improvement of thermopower.

**Acknowledgements**

We are grateful for the experimental assistance provided by Dr. Julio Martinez, John Nogan, Anthony R. James, Douglas V. Pete, Denise B. Webb, Renjie Chen and Wencao Yang. We acknowledge support, in part, from LDRD program at Los Alamos National Laboratory, start-up funding at Indiana University and NSF grant via DMR-1506460. We thank the Nanoscale Characterization Facility at Indiana University for instrumentation access. The XPS instrument was funded by NSF Award DMR MRI-1126394. This work was performed, in part, at the Center for Integrated Nanotechnologies, an Office of Science User Facility operated for the U.S. Department of Energy (DOE) Office of Science by Los Alamos National Laboratory (Contract DE-AC52- 06NA25396) and Sandia National Laboratories (Contract DE-AC04-94AL85000).

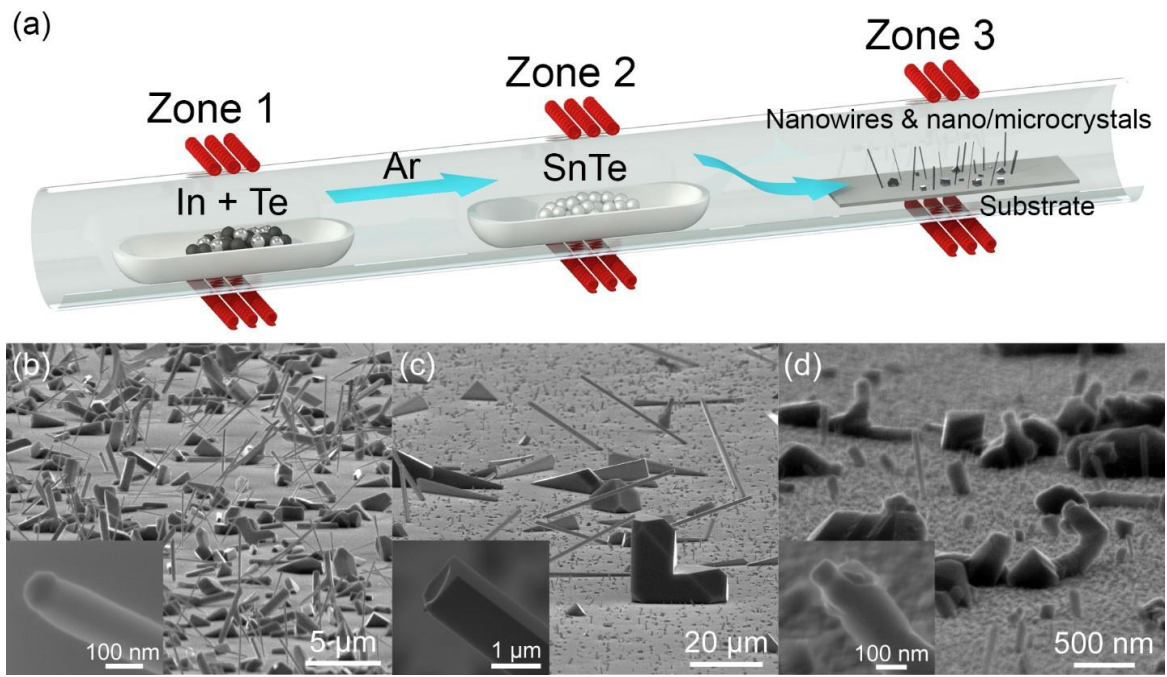

Figure 1 (a) Schematic picture of the chemical vapor deposition system used for the growth of nanostructures. The small quartz tubes that hold individual crucibles and substrates are not shown in the picture. SEM images taken on the (b) first substrate (closest to the precursors), (c) second substrate (further away from the precursors) and (d) third substrate (farthest from the precursors). Insets are magnified SEM images on the tip of a typical nanowire from (b) first substrate, (c) second substrate and (d) third substrate respectively.

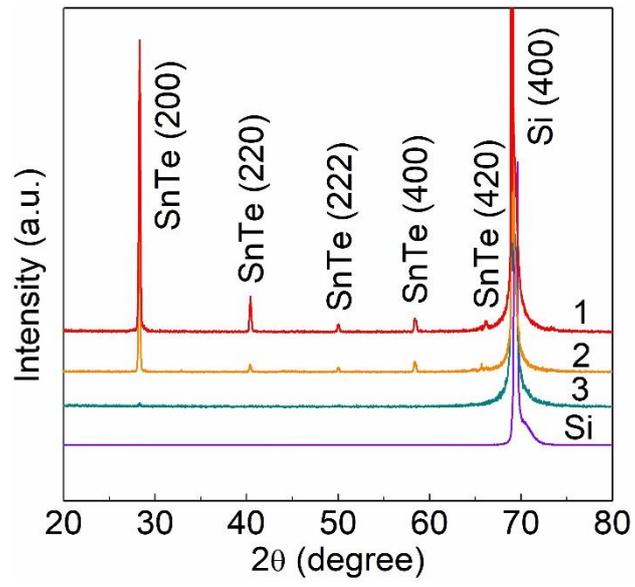

**Figure 2** XRD $\theta - 2\theta$ scans of nanostructures on the first, second, and third substrates as well as of a bare silicon substrate.

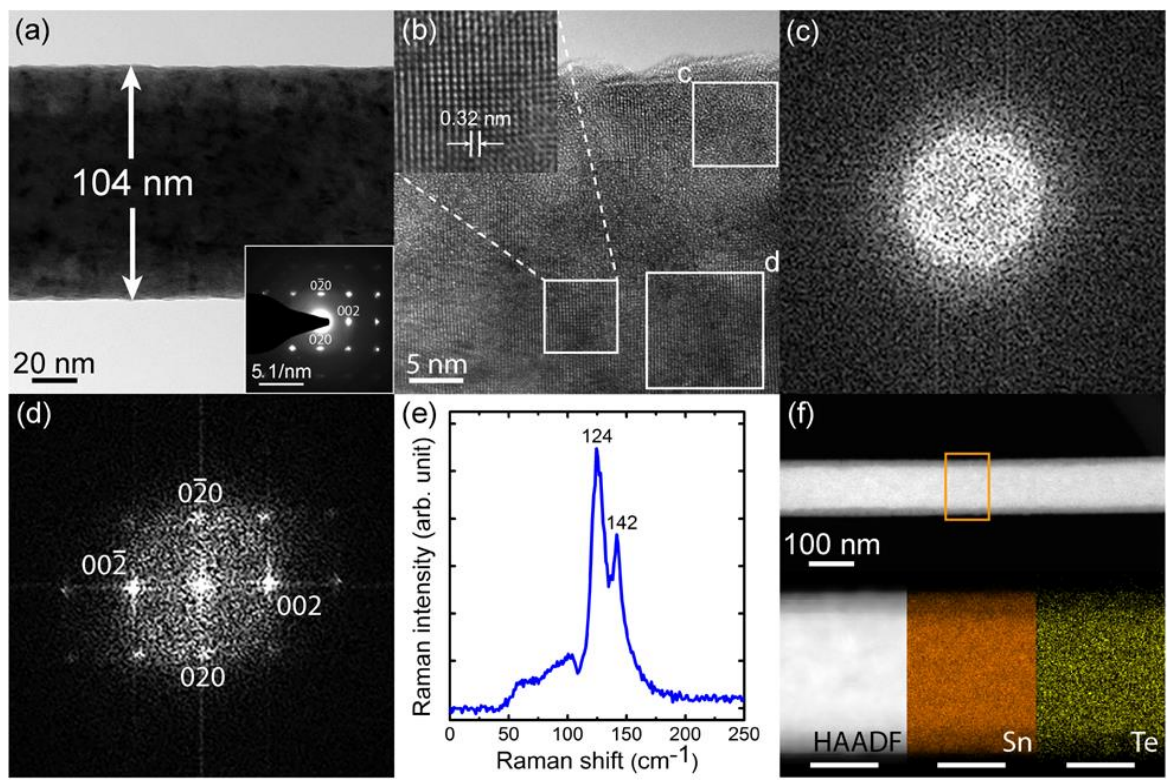

Figure 3 (a) Low resolution TEM image of a typical In-doped SnTe nanowire after prolonged exposure to air. Inset is the SAED pattern taken from the same nanowire. (b) HRTEM image of the nanowire body. FFT filtering was applied to manifest the lattice fringes. Two white frames indicate the location where FFT was conducted. (c) FFT taken near the edge of the nanowire. (d) FFT taken in the core of the nanowire. (e) Raman spectrum from a typical nanowire. (f) HAADF-STEM image of another nanowire and the EDX mapping showing individual elements. The scale bars are 50 nm.

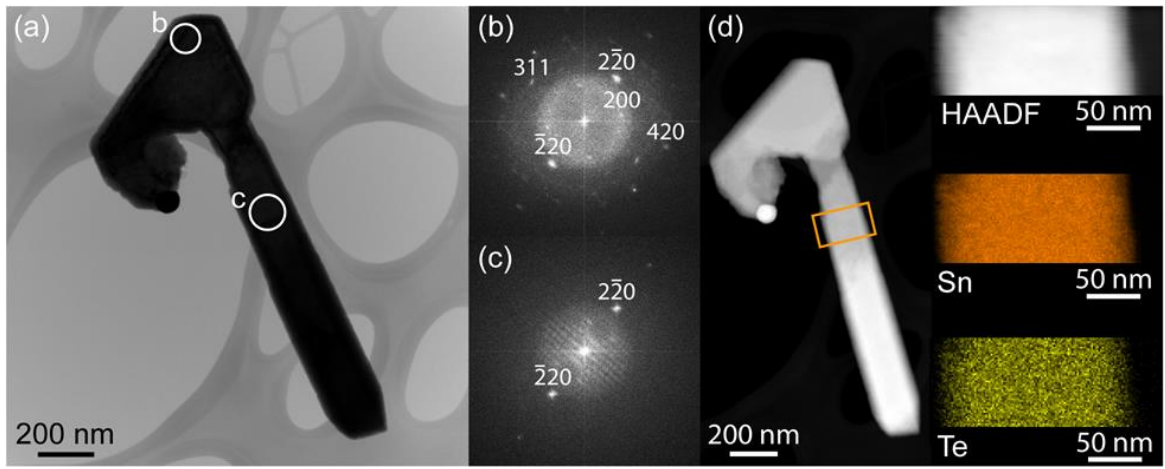

Figure 4 (a) Low resolution TEM image of a typical nanowire from the third substrate. (b) and (c) are FFT taken from the circled areas in (a). (c) HAADF-STEM image of the nanowire and EDX mapping showing individual elements.

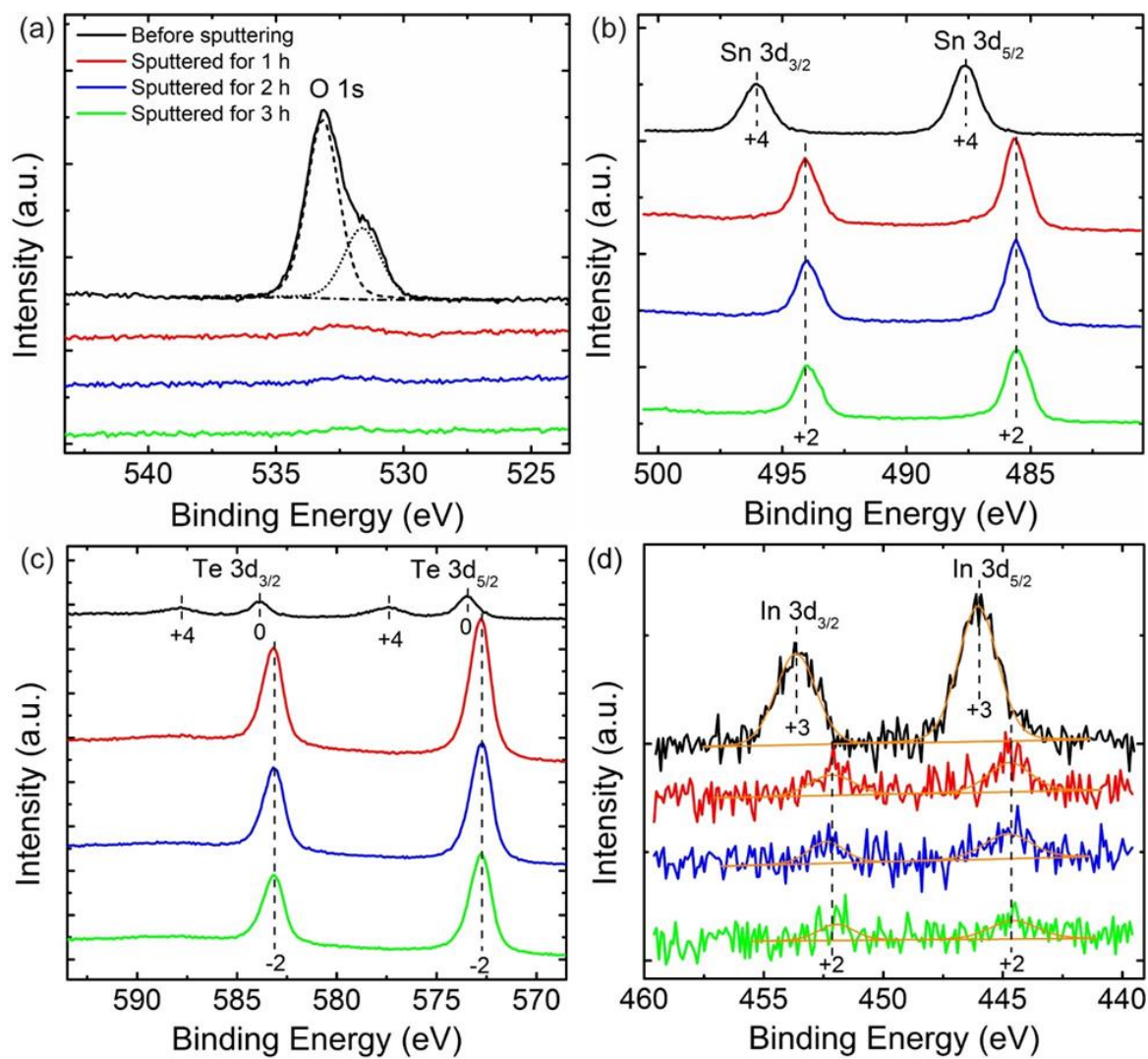

Figure 5 XPS spectra: (a) O 1s, (b) Sn 3d, (c) Te 3d and (d) In 3d core level spectra taken immediately before sputtering and after 1, 2 and 3 hours of sputtering respectively. All spectra were calibrated using highly oriented pyrolytic graphite (HOPG) due to the absence of carbon 1s peak after sputtering. For the spectra taken before sputtering, calibration using carbon 1s peak yields slightly lower binding energies. The dashed lines indicate roughly the peaks corresponding to different oxidation states. The spectra are vertically shifted for clarity.

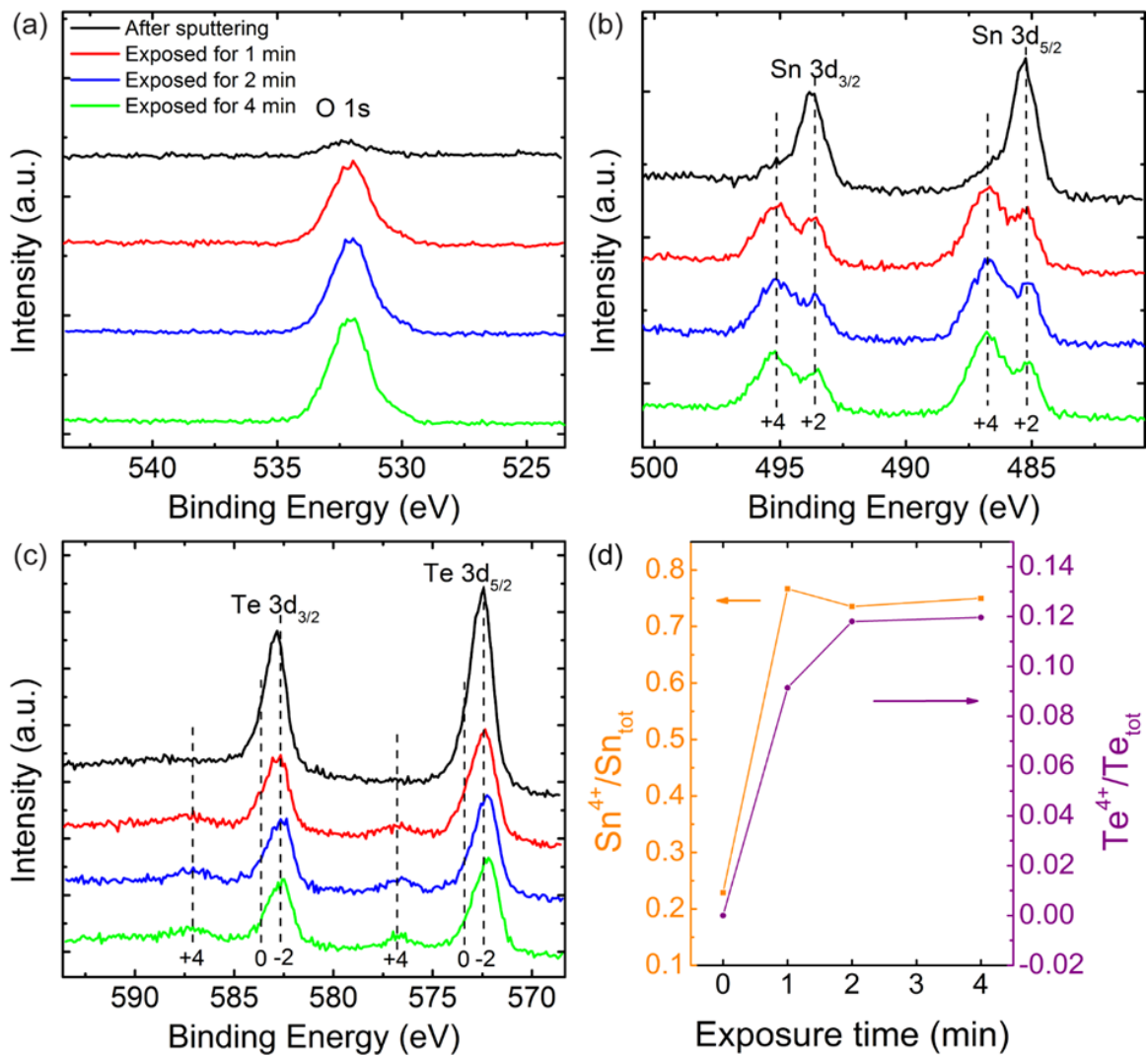

Figure 6 XPS spectra: (a) O 1s, (b) Sn 3d and (c) Te 3d core level spectra taken after sputtering, after 1, 2 and 4 minutes of exposure in atmosphere respectively. The spectra are vertically shifted for clarity. The dashed lines indicate roughly the peaks corresponding to different oxidation states. (d) Atomic ratio of $Sn^{4+}/Sn_{total}$ and $Te^{4+}/Te_{total}$ as a function of exposure time.

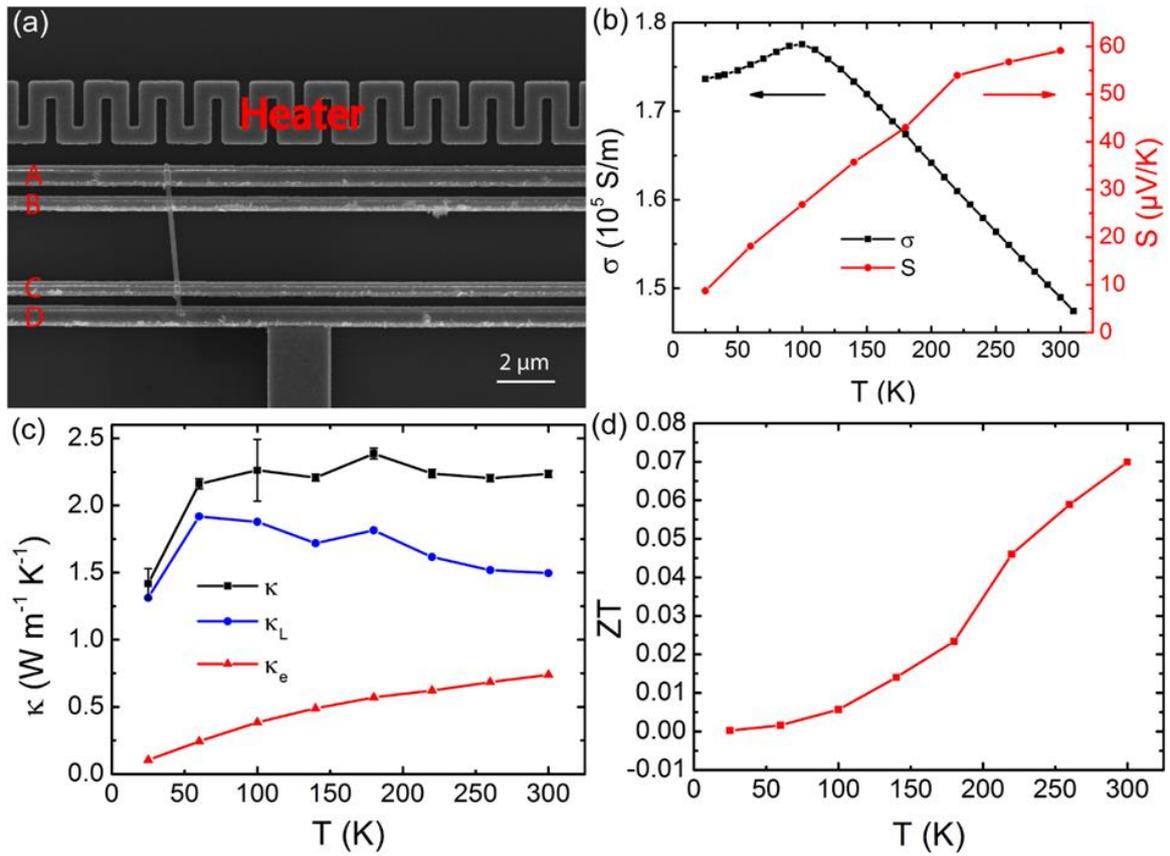

**Figure 7** (a) SEM image of an In-doped SnTe device for thermoelectric measurement. (b) Electrical conductivity and thermopower as a function of temperature. (c) Total thermal conductivity κ, electronic thermal conductivity $κ_e$ and lattice thermal conductivity $κ_L$ versus temperature. (d) ZT plotted against temperature.

**Table 1** Sample compositions after sputtering treatment and re-oxidation in atmosphere

| Sputtering treatment | In / Sn / Te | Re-oxidation | In / Sn / Te |
|---|---|---|---|
| Before sputtering | 3.3/77.4/19.3 | After sputtering | 0.7/58.7/40.6 |
| Sputtered for 1 h | 0.6/55.1/44.3 | Exposed for 1 min | 0.8/66.7/32.5 |
| Sputtered for 2 h | 0.7/56.0/43.3 | Exposed for 2 min | 0.6/68.8/30.7 |
| Sputtered for 3 h | 0.4/56.7/42.8 | Exposed for 4 min | 0.6/67.9/31.6 |